\documentclass[12pt]{iopart}

\usepackage[final]{epsfig}

\begin{document}

\title{Parton-medium interaction from RHIC to LHC --- a systematic approach}

\author{T Renk}

\address{Department of Physics, P.O. Box 35, FI-40014 University of Jyv\"askyl\"a, Finland}
\ead{thorsten.i.renk@jyu.fi}

\begin{abstract}
Despite a wealth of experimental data for high $p_T$ processes in heavy-ion collisions, discriminating between different models of hard parton-medium interactions has been difficult. One important reason is that the pQCD parton spectrum at RHIC kinematics is so steeply falling that distinguishing even a moderate shift in parton energy from complete parton absorption is almost impossible in observable quantities. In essence, energy loss models are effectively only probed in the vicinity of zero energy loss and as a result, only the pathlength dependence of energy loss offers some discriminating power at RHIC kinematics. At LHC, this is no longer the case: Due to the much flatter shape of the parton spectra originating from 2.76 AGeV collisions, the available observables probe much deeper into the model dynamics. A simultaneous fit of the nuclear suppression both at RHIC and LHC kinematics has thus a huge potential to discriminate between various models with equally good description of RHIC data alone.
\end{abstract}

\section{Introduction}

The suppression of the high transverse momentum $P_T$ hadron yield in heavy-ion (A-A) collisions as compared to the scaled expectation from p-p collisions, often referred to as 'jet quenching', has long been considered one of the most important probes of the medium created in heavy-ion collisions \cite{Jet1,Jet2,Jet3,Jet4,Jet5,Jet6}. Yet, despite a wealth of experimental data and several years of theoretical efforts, even solid qualitative statements such as to the nature of parton-medium interaction remain elusive. 

Several reasons contribute to the problem. First, there is an inherent ambiguity between modelling the parton-medium interaction and modelling the spacetime evolution of the underlying medium: an increased medium density or spatial size can usually compensate for a decrease of interaction strength. Second, primary parton spectra in perturbative Quantum Chromodynamics (pQCD) are steeply falling functions of $p_T$, and thus even a moderate shift in parton momentum is indistinguishable from an absorption of a parton by the medium \cite{gamma-h}. Finally, the problem itself is a genuinely difficult one, involving the hard scales of the perturbative process, the soft scales of medium dynamics and the transition region.

In this work, we show that a systematic comparison of combinations of medium evolution and parton-medium interaction models is capable of resolving the inherent ambiguities to a large degree, whereas the increased kinematic reach of the LHC as compared to RHIC significantly increases the dependence of observable quantities on model details.

\section{Modelling outline}

We start by selecting both a model for the medium spacetime evolution as well as a model for the parton-medium interaction. The spacetime evolution for RHIC conditions is chosen out of a range of fluid dynamical models which are constrained by bulk observables, a 3+1d ideal hydrodynamical model \cite{hyd3d}, a 2+1d ideal model \cite{hyd2d} and a viscous hydrodynamical code \cite{vhyd} (the latter with both a Glauber (vGlb) and a CGC-type (vCGC) initial condition).

Hard interactions are assumed to take place inside this evolving medium distributed in the transverse plane with a binary collision profile. From a given collision vertex, partons are propagated outward and undergo interactions with the medium. Here, we consider two main classes of parton-medium interaction models: full in-medium showers and leading parton energy loss models. In an in-medium shower evolution model, the yield of high $P_T$ hadrons can be computed from the expression
\begin{equation}
\label{E-Conv}
d\sigma_{\rm med}^{AA\rightarrow h+X}  = \sum_f d\sigma_{vac}^{AA \rightarrow f +X} \otimes \langle D_{MM}^{f \rightarrow h}(z,\mu^2)\rangle_{T_{AA}}
\end{equation} 
where $f$ sums over all parton flavours, $ d\sigma_{vac}^{AA \rightarrow f +X}$ is the vacuum pQCD cross section for producing parton $f$ and $\langle D_{MM}^{f \rightarrow h}(z,\mu^2)\rangle_{T_{AA}}$ is the geometry-averaged medium modified fragmentation function (MMFF) for fractional momentum $z$ at scale $\mu^2$.  The MMFF is the output of a parton-medium interaction model given the path through the medium. Here we use the Monte Carlo (MC) code YaJEM \cite{YaJEM1,YaJEM2,YaJEM-D} to compute it. The geometry averaging is done over all possible initial vertices, either with a given orientation with respect to the event plane or averaged over all orientations.
 In leading parton energy loss models, the MMFF is approximated by
\begin{equation}
\langle D_{MM}^{f \rightarrow h}(z,\mu^2)\rangle_{T_{AA}} = \langle P(\Delta E)\rangle_{T_{AA}} \otimes D^{f \rightarrow h}(z, \mu^2)
\end{equation}
i.e. by a convolution of the vacuum fragmentation function $D^{f \rightarrow h}(z, \mu^2)$ with a geometry-averaged energy loss probability distribution $\langle P(\Delta E)\rangle_{T_{AA}}$. This latter quantity is computed within a given leading parton energy loss framework. In the present study we consider a radiative energy loss model \cite{ASW} (ASW), a parametrized elastic energy loss model \cite{Elastic} (elastic) as well as a MC model for elastic pQCD interactions \cite{ElasticMC} (eMC) and an AdS/CFT inspired model for energy loss in a strongly coupled medium \cite{AdS} (AdS).

In each of these models, a single parameter $K_{med}$ regularizes the proportionality between powers of thermodynamical quantities such as energy density $\epsilon$ or temperature $T$ and the interaction strength. Unless stated otherwise, we adjust $K_{med}$ for any combination of medium evolution and parton-medium interaction model such that the nuclear suppression factor $R_{AA}$ in central 200 AGeV Au-Au collisions at RHIC is reproduced and compute for other centralities, different $\sqrt{s}$ or other observables without additional free parameters.

\section{Pathlength dependence}

Different parton-medium interaction models show different response to the pathlength $L$ of a parton propagating in a constant medium. This can be exploited to discriminate between models. For instance, any incoherent process (e.g. elastic or eMC) counts the number of interactions along the path by $n_{scatt} = L/\lambda$ where $\lambda$ is the mean free path. If the energy loss of a parton to the medium is proportional to $n_{scatt}$, a linear pathlength dependence of the total lost energy $\Delta E \sim L$ follows. On the other hand, in coherent radiative processes, the virtuality $Q$ of a gluon with energy $\omega$ from the virtual cloud surrounding a parton must be brought on-shell by random transverse kicks from the medium. While the number of kicks is proportional to $L$, there is also a coherence condition which states that interactions within the formation time $\tau \sim \omega/Q^2$ need to be summed coherently. This implies a quadratic pathlength dependence $\Delta E \sim L^2$ \cite{ASW} (ASW). If however in addition finite energy corrections are accounted for, such a quadratic dependence effectively reverts back to a linear dependence for experimentally relevant kinematics \cite{YaJEM-D} (YaJEM). In a strongly coupled medium where an AdS/CFT description may be applicable, the gluons from the virtual cloud surrounding the parent parton are not brought on-shell by random transverse kicks but by the action of a drag force of order $T^2$. Coherence time arguments in this case lead to $\Delta E \sim L^3$ \cite{AdS} (AdS). Finally, an in-medium shower corresponds to the evolution from a high initial virtuality scale down to a low scale $Q_0$. If one takes into account that the medium can only affect a shower above $Q_{med} = \sqrt{E/L}$ where $E$ is the parent parton energy, an explicit non-linear behaviour of the medium effect with pathlength and energy $E$ emerges \cite{YaJEM-D} (YaJEM-D).

Thus, the different physics assumptions underlying various models are reflected in the expected dependence of medium modification on pathlength. Experimentally this is accessible e.g. through the emission of high $P_T$ hadrons as a function of the angle $\phi$ with the event plane, or specifically in the difference of in-plane and out of plane emission.  However, while interesting to characterize models, pathlength dependence in a constant medium is not relevant for the experimental situation. In a real hydrodynamical evolution, the spatial density profile, longitudinal and transversal flow, viscous reheating and fluctuations in the initial state all have noticeable influence \cite{JetHydSys,JetFluct}, underlining the need for realistic hydrodynamically modelling and an assessment of the uncertainties associated with the medium model.

\begin{figure}[htb]
\epsfig{file=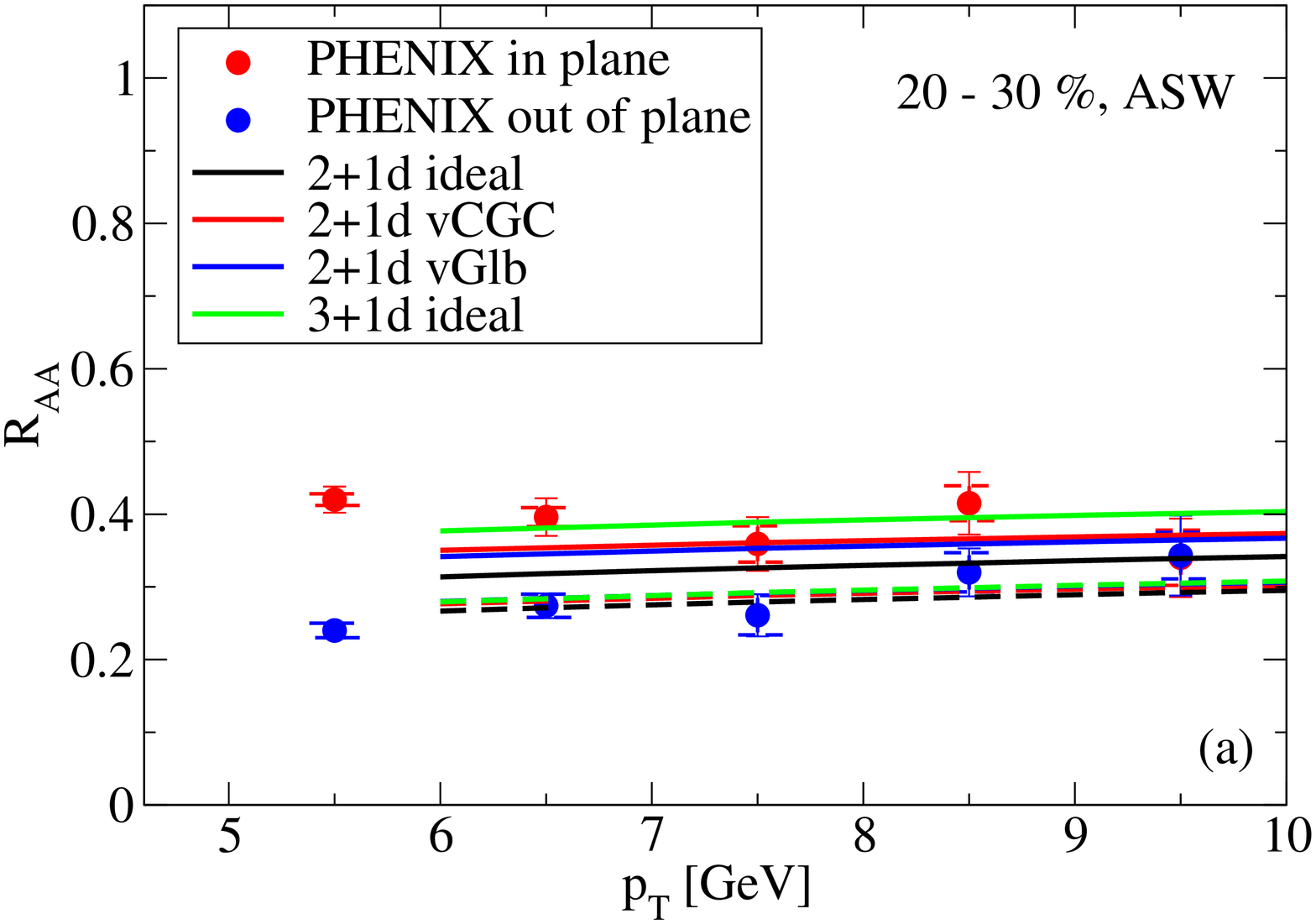, width=7.8cm}\epsfig{file=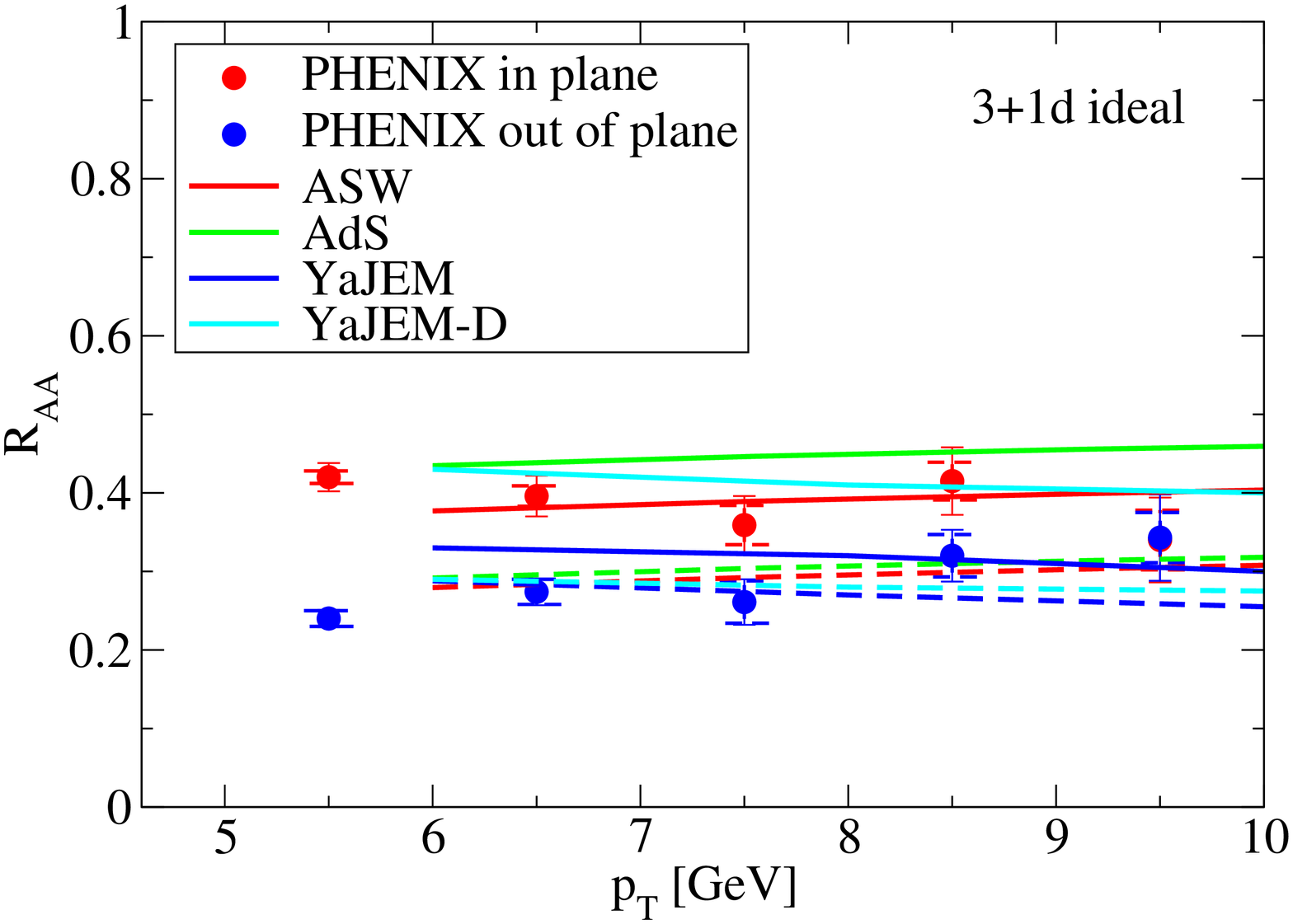, width=7.8cm}
\caption{\label{F-Pathlength}Left panel: $R_{AA}(P_T)$ for 30-40\% cental 200 AGeV Au-Au collisions for in plane (solid) and out of plane (dashed) emission computed for the same energy loss model (ASW) with different hydrodynamical backgrounds, compared with PHENIX data \cite{PHENIX_RAA_phi}. Right: same as left panel, but for the same 3+1d ideal hydrodynamical model and different parton-medium interaction models.}
\end{figure}

In Fig.~\ref{F-Pathlength} we show some results of a systematic investigation \cite{JetHydSys} of pathlength dependent observables for different combinations of medium evolution and parton-medium interaction model. Clearly, both elements have a pronounced influence on the results. Summarizing the findings of \cite{JetHydSys}, we can state that the spread between in plane and out of plane emission grows whenever energy loss happens late. This may be due to the $L^3$ dependence of AdS or the strong non-linear dependence of YaJEM-D, but can also be driven by the hydrodynamical component. Here, differences are unrelated to the dimensionality of the simulation, rather factors like initialization time, equation of state, viscosity or freeze-out conditions matter. There is no single factor which influences the spread, rather a combination of various effects contributes with almost equal magnitude. 

Several combinations of models are viable, however for instance an $L$ dependence as characteristic for incoherent interactions or of radiative energy loss with finite energy corrections fails no matter what medium is assumed \cite{YaJEM-D,Elastic,ElasticMC}. Other models work conditionally, for instance the AdS model works fine with the 2+1d hydrodynamics but overestimates the spread seen in the data for the 3+1d model, whereas the ASW model behaves the opposite way. Thus, while pathlength dependent observables clearly have some power to distinguish various scenarios of parton-medium interaction and/or do medium tomography, additional constraints are needed.

\section{Hydrodynamics with fluctuating initial conditions}

Before assessing the potential of a larger kinematic lever-arm to discriminate in more detail between various model calculations, let us discuss a potentially troublesome issue connected with modelling the medium. It has recently become apparent that event-by-event fluctuations in the initial conditions are crucial to understand details of the hydrodynamical evolution of the medium in A-A collisions. In other words, it matters if the initial state is first averaged and then the evolution of an average final state is computed, or if the evolution for each initial state is computed and only the final state is averaged. One may thus wonder if the same is true for hard partons interacting with such a medium.
Potentially, there are several effects that might create a difference. First, $R_{AA}$ is a non-linear function of medium density which responds stronger to a decreasing density than to an increasing density (even an arbitrarily high density cannot push $R_{AA}$ below zero). Thus, fluctuations in the initial state may decrease the observed amount of suppression. However, 'hotspots' in the hydrodynamical initial state are typically associated with binary collision vertices. Taking this correlation into account implies that produced partons tend to be produced in regions with higher-than-average density, which would decrease the amount of suppression as compared to a smooth, initial-state averaged medium. In addition, there is also the effect of a very irregular initial flow field, for which the sign is {\itshape a priori} unknown. Finally, the fact that the event plane is not identical with the reaction plane needs to be taken seriously --- if the reaction plane is used as reference plane for $R_{AA}(\phi)$, then the magnitude of the spread between in-plane and out of plane emission is artificially  decreased by a trivial averaging effect.

\begin{figure}[htb]
\epsfig{file=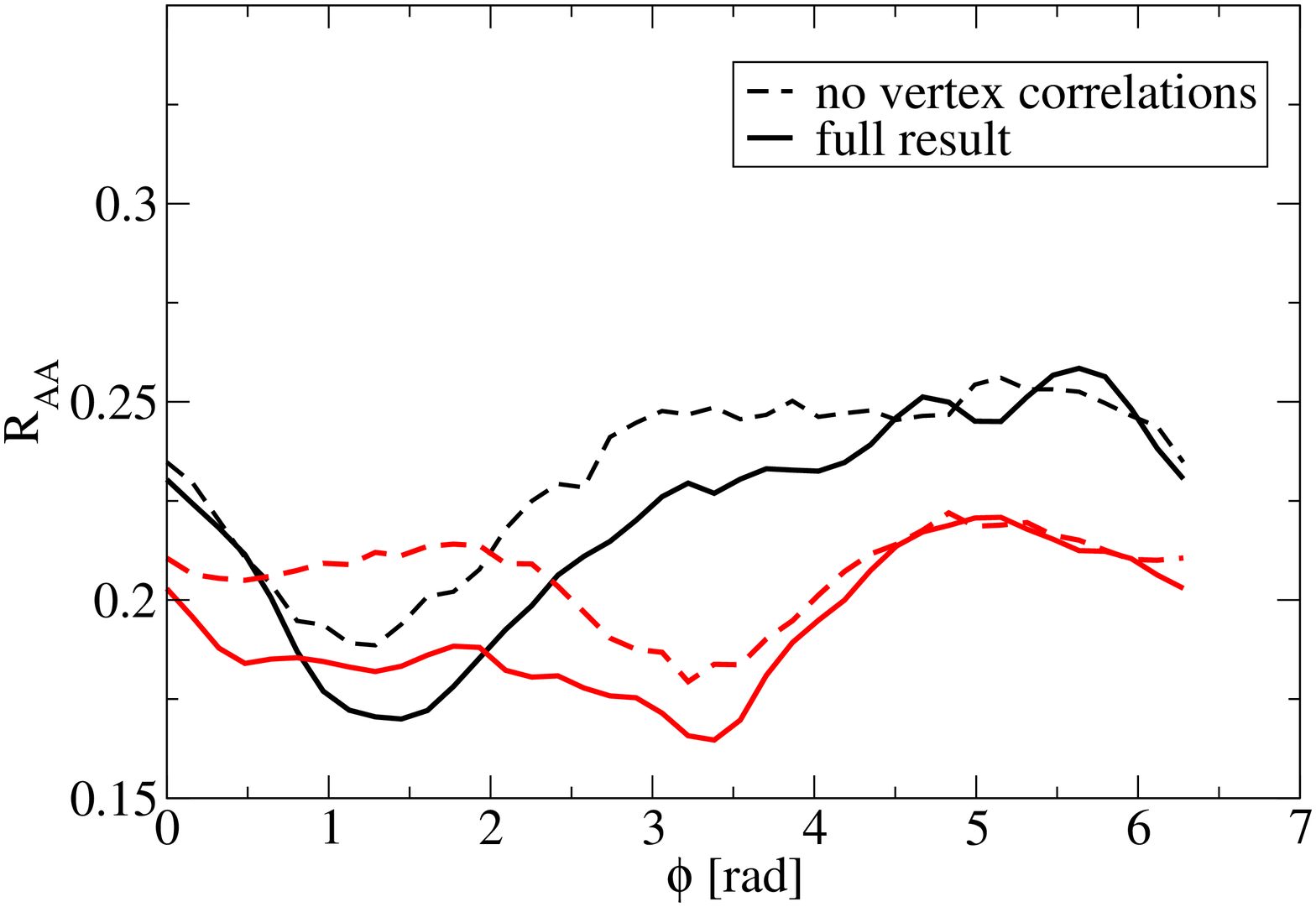, width=7.8cm}\epsfig{file=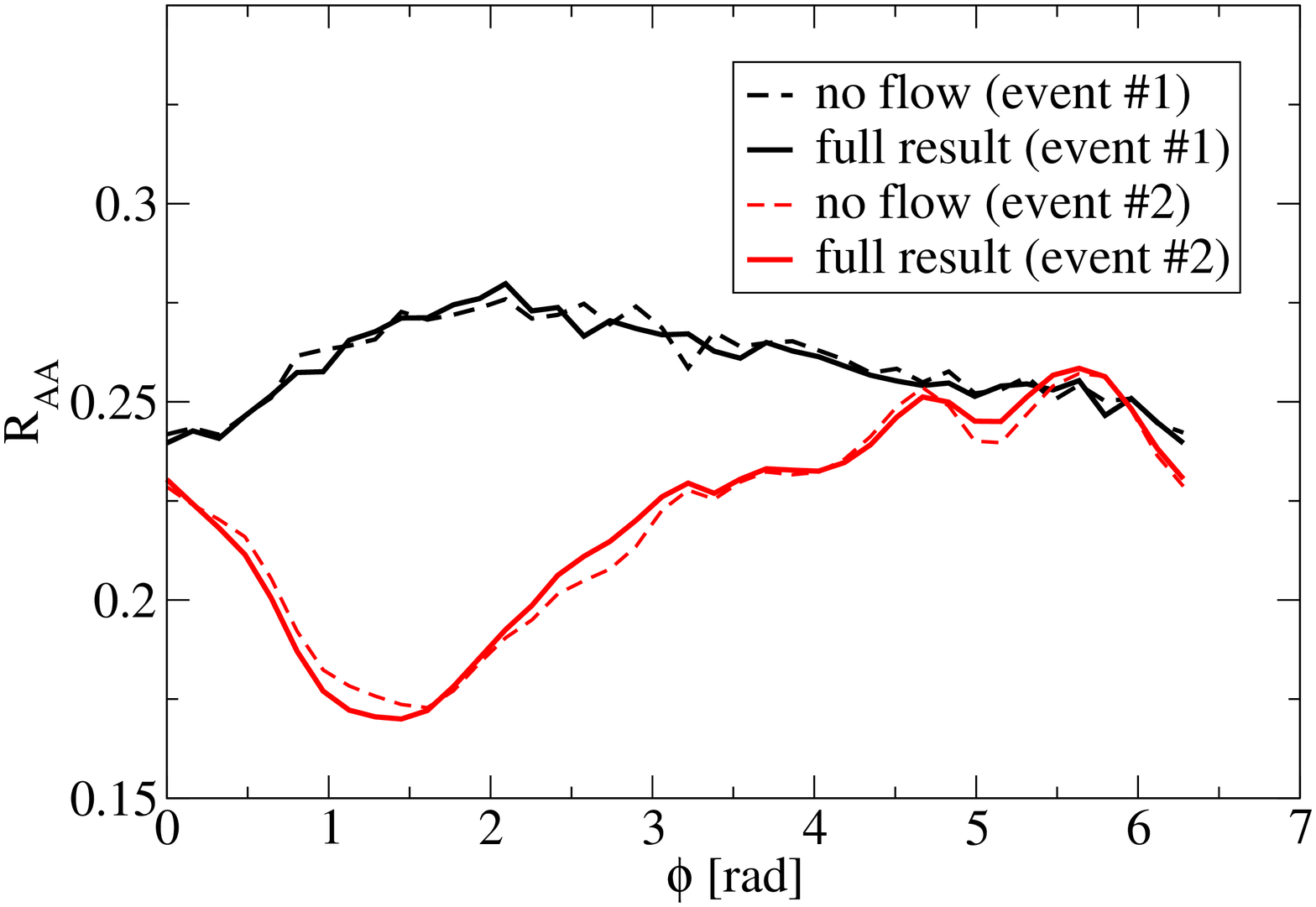, width=7.8cm}
\caption{\label{F-Fluct}Left: $R_{AA}(\phi)$ computed in the ASW model for two different events, with (solid) and without (dashed) taking the correlation between binary collisions and hotspots into account. Right: same as left, but with (solid) and without (dashed) taking the irregular fluctuation-driven flow field into account.}
\end{figure}

In order to investigate the role of fluctuations, we use the ASW energy loss model in combination with a 2+1d ideal hydrodynamical model with initial state fluctuations and study $R_{AA}(\phi)$ at fixed $P_T = 10$ GeV \cite{JetFluct}.  In Fig.~\ref{F-Fluct} we show for two events each the effects of the correlation of the production vertex with hotspots and of the irregular flow field.
We find that both inter- and intra-event fluctuations are sizeable. The net effect of the correlation is indeed a downward shift of $R_{AA}$ as expected, whereas the effect of the fluctuation-driven flow field is very mild. In central events the cancellation between nonlinearity and vertex-hotspot correlation is very good. When we perform a 20 event average and aim for a best fit to the data we obtain a $K_{med}$ which is less than 20\% different from the smooth case. In non-central collisions, the cancellation becomes imperfect and dependent on the size scale of the fluctuations, resulting in a decrease of suppression for small-scale fluctuations. This can potentially be used to constrain the physics origin of fluctuations. Our study would indicate relatively large-scale O(0.8) fm fluctuations. All in all, the observed magnitude of effects does not suggest that fluctuations in the hydrodynamical initial state are a sizeable effect for high $P_T$ probes, thus conclusions obtained using smooth hydrodynamical models remain essentially valid.
Qualitatively similar results have also been obtained with the eMC model \cite{ElasticMC}.

\section{$P_T$ dependence of $R_{AA}$ at larger $\sqrt{s}$}

As mentioned initially, for a steeply falling parton spectrum even small shifts in parton momentum lead to a large suppression in hadron yield, which makes observables at RHIC insensitive to details of the parton-medium interaction \cite{gamma-h}. At LHC kinematics with 2.76 ATeV, this is no longer the case since the spectra are much harder. Thus, the $P_T$ dependence of $R_{AA}$ is now not only visible, but carries information about model details. 
However, in order to take advantage of this fact, one needs to overcome the ambiguity due to the modelling of the soft medium evolution. Ideally, one would like to run 'the same' hydro at larger $\sqrt{s}$ in order to connect with RHIC results, but  in practice a hydro code does not take $\sqrt{s}$ as input parameter but rather an initial condition in terms of entropy distribution and thermalization time and a breakup condition. Thus, additional modelling is required to constrain the $\sqrt{s}$ dependence of these quantities.

\begin{figure}[htb]
\epsfig{file=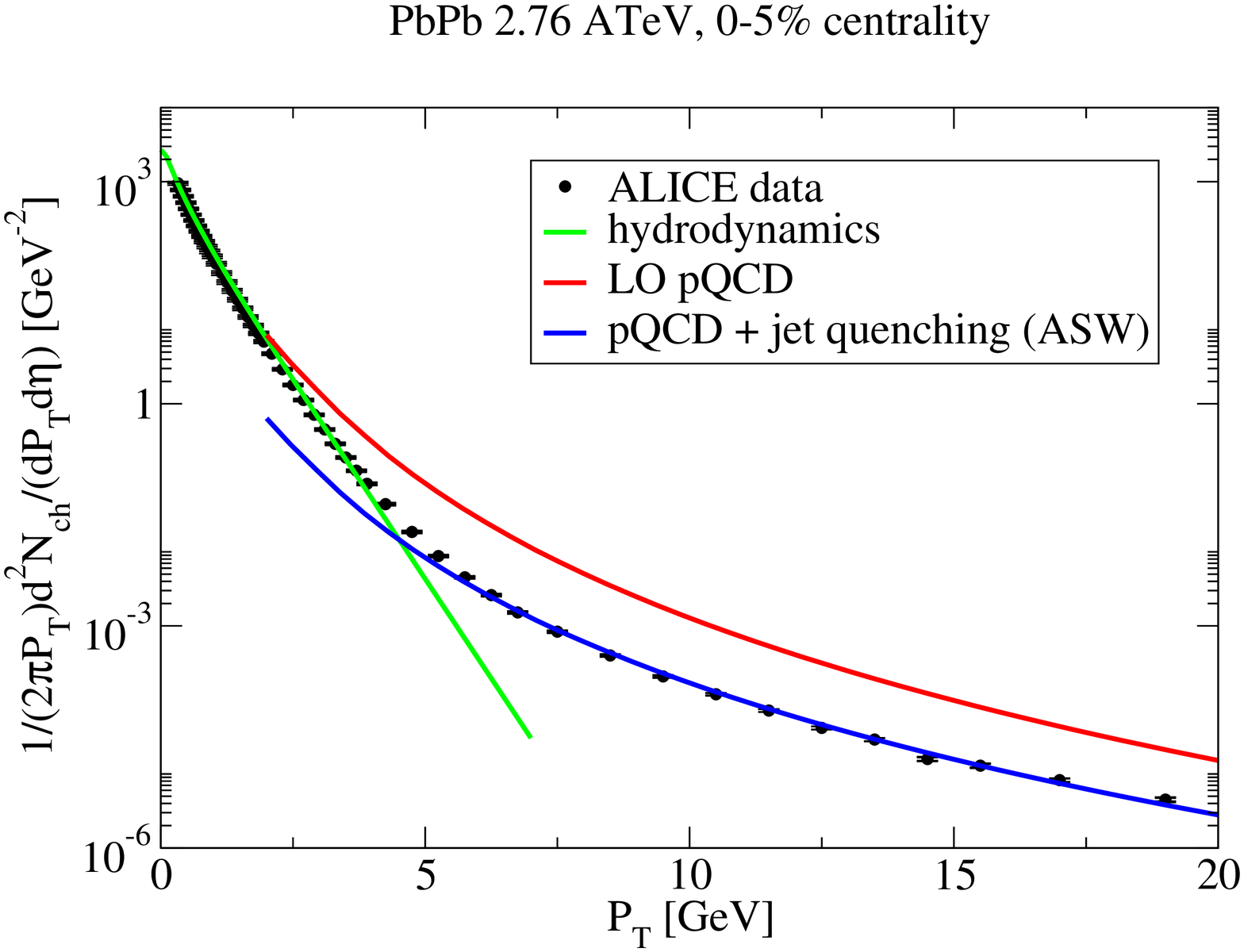, width=7.9cm}\epsfig{file=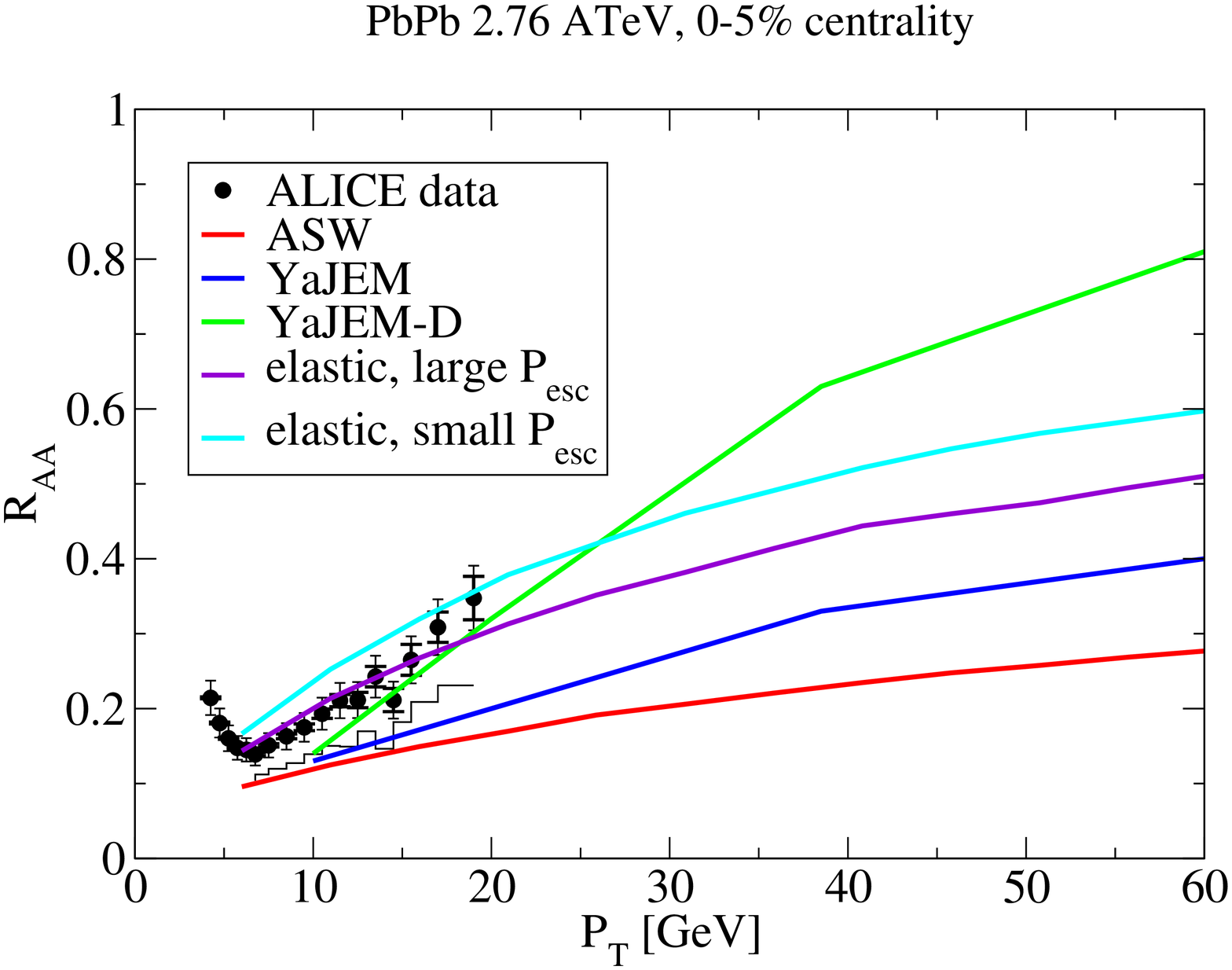, width=7.7cm}
\caption{\label{F-LHC}}
\end{figure}

Here, we use the EKRT initial state saturation model \cite{EKRT} and results from a dynamical freeze-out criterion \cite{DynFo} to constrain the extrapolation of the medium evolution from RHIC to LHC \cite{RAA_LHC}. In Fig.~\ref{F-LHC} left, we demonstrate that this procedure combined with a pQCD + parton-medium interaction component can give good agreement throughout the whole $P_T$ range measured by the ALICE collaboration \cite{ALICE}. In Fig.~\ref{F-LHC} right, we show that, as expected, the various parton medium interaction models tuned to RHIC data predict very different results at LHC kinematics. While currently a large systematic uncertainty in the measurement prevents any firm conclusions, this shows, in combination with pathlength dependent observables, the potential to uncover that nature of the parton-medium interaction. 

\section{Conclusions}

While fully reconstructed jets are the observable most closely reflecting the QCD dynamics of parton evolution in the medium, they also suffer from conceptual problems separating jet from medium at soft scales. In contrast, the study presented here demonstrates that a systematic investigation of model combinations against a significant body of single inclusive high $P_T$ hadron data (as well as correlation) has the potential to identify or at least significantly constrain the physics of parton-medium interaction without running into scale sepataion problems. Future high-precision data from the LHC experiments will therefore quickly identify viable models for jet productions in medium if the constraints from leading hadron and hadron correlation physics are taken seriously.

\section*{Acknowledgements}

Fruitful collabortion with K.~J.~Eskola, H.~Holopainen, J.~Auvinen, R.~Paatelainen, C.~Marquet, U.~Heinz and C.~Shen is gratefully acknowledged.  This work is 
supported by the Academy Researcher program of the Finnish Academy (Project 130472).

\end{document}